\documentclass[a4paper]{jpconf}
\usepackage{graphicx}
\usepackage{amsmath,amssymb}
\usepackage{hyperref} 

\usepackage{color}

\begin{document}
%%%%%%%%%%%%%%%%%%%%%%%%%%%%%%%%%%%%%%%%%%%%%%%%%%
\title{Fidelity, susceptibility and critical exponents in the Dicke model}
\author{M. A. Bastarrachea-Magnani, O. Casta\~nos, E. Nahmad-Achar, R. L\'opez-Pe\~na, and J. G. Hirsch}
\address{Instituto de Ciencias Nucleares, Universidad Nacional Aut\'onoma de M\'exico \\ Apdo. Postal 70-543, M\'exico D. F., C.P. 04510}
\ead{miguel.bastarrachea@nucleares.unam.mx}

%%%%%%%%%%%%%%%%%%%%%%%%%%%%%%%%%%%%%%%%%%%%%%%%%%

%%%%%%%%%%%%%%%%%%%%%%%%%%%%%%%%%%%%%%%%%%%%%%%%%%
\begin{abstract}
We calculate numerically the fidelity and its susceptibility for the ground state of the Dicke model. A minimum in the fidelity identifies the critical value of the interaction where a quantum phase crossover, the precursor of a phase transition for finite number of atoms $\mathcal{N}$, takes place. The evolution of these observables is studied as a function of $\mathcal{N}$, and their critical exponents evaluated. Using the critical exponents the universal curve for the specific susceptibility is recovered. An estimate to the precision to which the ground state wave function is numerically calculated is given, and found to have its lowest value, for a fixed truncation, in a vicinity of the critical coupling.  
\end{abstract}
%%%%%%%%%%%%%%%%%%%%%%%%%%%%%%%%%%%%%%%%%%%%%%%%%%

\section{Introduction}
The Dicke Hamlitonian describes a system of $\mathcal{N}$ two-level atoms interacting with a single monochromatic electromagnetic radiation mode within a cavity \cite{Dicke54}. In terms of quantum computation, it can also describe a set of $\mathcal{N}$ qubits, realized with quantum dots, Bose-Einstein condensates or QED circuits \cite{Sche07}, interacting through a bosonic field. In recent years Dicke-like Hamiltonians, and in particular its quantum phase transition (QPT) from normal to superradiant behavior \cite{Hepp73,Wang73}, have attracted much attention. The QPT is an example of a quantum collective behavior and has a close connection with entanglement and quantum chaos. Besides, the Dicke Hamiltonian for a finite $\mathcal{N}$ provides a good description for the systems manipulated in the laboratory, especially in the light of the experimental realization of the superradiant phase transition in a BEC \cite{Bau10,Nag10}, the intense development in the control of single atoms and photons in a cavity, and the possibility of a QPT in a system of  $\mathcal{N}$ QED circuits \cite{Vieh11,Cuiti12}.  
 
The Dicke model can be written as ($\hbar=1$)
\begin{equation}
H_{D}=\omega a^{\dagger}a+\omega_{0}J_{z}+\frac{\gamma}{\sqrt{\mathcal{N}}}\left(a+a^{\dagger}\right)\left(J_{+}+J_{-}\right) ,
\end{equation}
where $\omega$ is the field frequency, $\omega_{0}$ is the atomic energy separation, $a^{\dagger}$ and $a$ are the creation and annihilation photon operators, respectively, and $\gamma$ is the coupling strength. $J_{z}$ and $J_{\pm}$ are collective atomic operators (pseudospin operators) which follow the SU(2) algebra, and denote the atomic relative population and the atomic transitions operators, respectively. 

For a finite number of atoms $\mathcal{N}$, the model is in general non-integrable, and care must be taken when the first order in the $1/\mathcal{N}$ expansion is employed because of its singular behavior around the phase transition \cite{OCasta11a,OCasta11}. The Hamiltonian is integrable in at least two limits ($\gamma\rightarrow 0$ and $\omega_{0}\rightarrow 0$). In the thermodynamic limit, when the number of atoms $\mathcal{N}$ goes to infinity, the mean field description becomes exact. It provides analytic asymptotic solutions through a Holstein-Primakoff expansion \cite{Emary03}, which allows to extract the critical exponents for the ground state energy per particle, the fraction of excited atoms, the number of photons per atom, their fluctuations and the concurrence \cite{Emary03,Lam05,Vid06,Chen08}. Numerical solutions complement and confirm the theoretical predictions, and allow for the exploration of the system in regimes which are not described by the latter, like excited state phase transitions. 

A concept emerging from quantum information theory, the fidelity, can be used to determine a sudden change in the ground state of a quantum system as a function of a control parameter. In recent years it has emerged as a powerful tool to study QPT in quantum many-body systems \cite{Gu10}. The fidelity describes the overlap between two quantum states. Considering a quantum many-body system, the general form of the Hamiltonian can be written as
\begin{equation}
H=H_{0}+\gamma H_{1} ,
\end{equation}
where $H_{1}$ is the interaction Hamiltonian and $\gamma$ is a control parameter. For two pure states $\left|\psi(\gamma)\right\rangle$ and $\left|\psi(\gamma')\right\rangle$ the fidelity is written as \cite{Gu10}
\begin{equation}
F(\gamma,\gamma')=|\left\langle\psi(\gamma)\right.\left|\psi(\gamma')\right\rangle|  .
\end{equation}  
The fidelity measures the amount of shared information between two quantum states, being its geometric interpretation the closeness of these states. Being a QPT a sudden change in the ground state properties of a system when a control parameter varies, a minimum in the fidelity allows to locate and characterize the QPT.  Its second derivative, the fidelity susceptibility, is even more sensitive to the QPT. Expanding the fidelity around its minimum, for  $\gamma-\gamma'$ small, we have \cite{Gu10}
\begin{equation}
F(\gamma,\gamma')=1-\frac{(\gamma-\gamma')^{2}}{2}\chi^{F}+... 
\end{equation} 
The fidelity susceptibility $\chi^{F}$ can be expressed as
\begin{equation}
\chi^{F}(\gamma)=\lim_{\gamma-\gamma'\rightarrow 0} \frac{-2\,\mbox{ln}F(\gamma,\gamma')}{(\gamma-\gamma')^{2}}=\frac{2(1-F(\gamma,\gamma'))}{(\gamma-\gamma')^{2}},
\end{equation} 
being the first form in terms of the logarithmic fidelity. It is useful to choose $\gamma'=\gamma+d\gamma$ in order to vary $\gamma$ while taking the limit $d\gamma \rightarrow 0$. 

In the thermodynamic limit, the fidelity goes to zero in the QPT, while the susceptibility goes to infinity. For finite systems, in the critical value of the coupling $\gamma_{max}$, the fidelity and its susceptibility show the \emph{precursor} of the QPT by obtaining a minimum and a maximum value, respectively. Calculating the behavior of  these quantities (the critical coupling parameter and the maximum value of the susceptibility) allows us to derive their critical exponents as a function of the number of atoms $\mathcal{N}$ \cite{Ocasta12, Nahmad12}. Furthermore, one can obtain universal curves for some observables like the fidelity \cite{Gu12} or the susceptibility. For a finite scale analysis, we can define a universal quantity called the specific susceptibility \cite{Kwok08},
\begin{equation}
\chi_{s}=\frac{\chi^{F}(\gamma_{max})-\chi^{F}(\gamma)}{\chi^{F}(\gamma)} .
\end{equation}
The specific susceptibility is useful to compare systems with different number of atoms. 

In this work we calculate the fidelity and its susceptibility for the ground state of the finite Dicke model, performing a numerical diagonalization of the Hamiltonian. Using the fidelity formalism we locate the precursor of the QPT for each $\mathcal{N}$. With it, we find numerically the critical exponent of the coupling parameter, which tends to $\gamma_{c}=\sqrt{\omega\omega_{0}}/2$, the critical value in the thermodynamic limit. We also study the behavior of the minimum of the fidelity and the maximum of its susceptibility as $\mathcal{N}$ grows, finding their critical exponents. We build the universal curve of the specific susceptibility, which confirms the value of the critical exponent. Finally, we make a brief discussion of the ground state wave function precision as a function of the coupling strength. 

%%%%%%%%%%%%%%%%%%%%%%%%%%%%%%%%%%%%%%%%%%%%%%%%%%

\section{Numerical solution} 
In order to solve numerically the Dicke Hamiltonian we employ extended bosonic coherent states \cite{Chen08,Basta11}. They are built with the displaced boson operators $A^{\dagger}, A$, obtained by shifting the original annihilation operator $a$:
\begin{equation}
A=a+\frac{2\gamma}{\omega\sqrt{\mathcal{N}}}J_{x} .
\end{equation}
The new basis is $\{ \left|N;j,m\right\rangle \}$, where $N$ is an eigenvalue of the new number operator $A^{\dagger}A$, $j= \mathcal{N}/2$ and $m$ is  an eigenvalue of $J_x$. It allows for the determination of ground state properties in the superradiant region far beyond previous attempts \cite{Chen08},  
and also of excited states with a single truncation \cite{Basta12}. 

To solve the Hamiltonian numerically we must truncate the Hilbert space, which is infinite due to the presence of the number operator in the Hamiltonian. 
In order to estimate the minimal truncation to be employed, we use a criterion based on the precision of the wave function, which we call the $\Delta P$ criterion \cite{Basta13}. We express the ground state wave function as:
\begin{equation}
|\Psi(N_{max})\rangle=\sum\limits_{N=0}^{N_{max}} \sum\limits_{m=-j}^{j} C_{N,m} |N;j,m\rangle,
\end{equation} 
where $C_{N,m}$ are the coefficients of the exact ground state wave function and $N_{max}$ is the value of the truncation in the number of displaced excitations. The probability $P_{N}$ of having $N$ excitations in the ground state is:
\begin{equation}
P_{N}=|\langle N|\Psi\rangle|^{2}=\sum_{m}|C_{N,m}|^{2}
\end{equation}
We define the precision in the calculated wave function as (see Appendix)
 \begin{equation}
 \Delta P=\sum\limits_{m=-j}^j \left|C_{N_{max}+1,m}(N_{max}+1)\right|^2.
 \end{equation}
By diagonalizing the Hamiltonian with several truncations, if $ \Delta P$ is smaller than certain tolerance we consider that the solution has converged, being $N_{max}$ the minimum value of the truncation necessary for obtaining the exact numerical solution. 

%%%%%%%%%%%%%%%%%%%%%%%%%%%%%%%%%%%%%%%%%%%%%%%%%%

\section{Results}
\begin{figure}
\begin{center}
\includegraphics[scale=0.5]{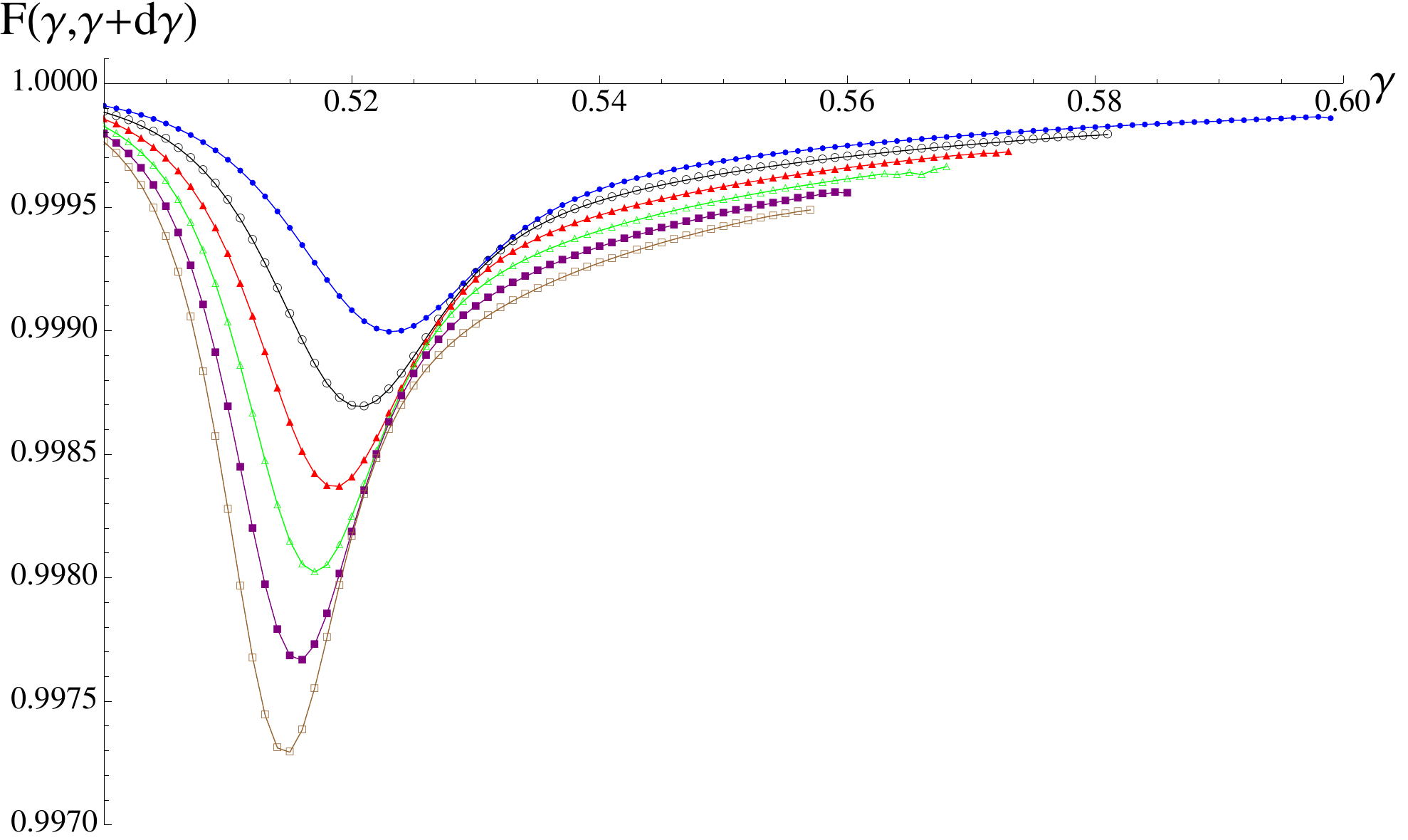}
\end{center}
\caption{\label{fig:1}Fidelity as a function of the coupling parameter. With $\gamma$ from $\gamma_{c}=0.5$ to $0.6$, $d\gamma=0.001$, $\omega_{0}=\omega=1$ and $\mathcal{N}=100$, $120$, $140$, $160$, $180$, $200$ (top to bottom). $N_{max}=8$.}
\end{figure}

\begin{figure}
\begin{center}
\includegraphics[scale=0.5]{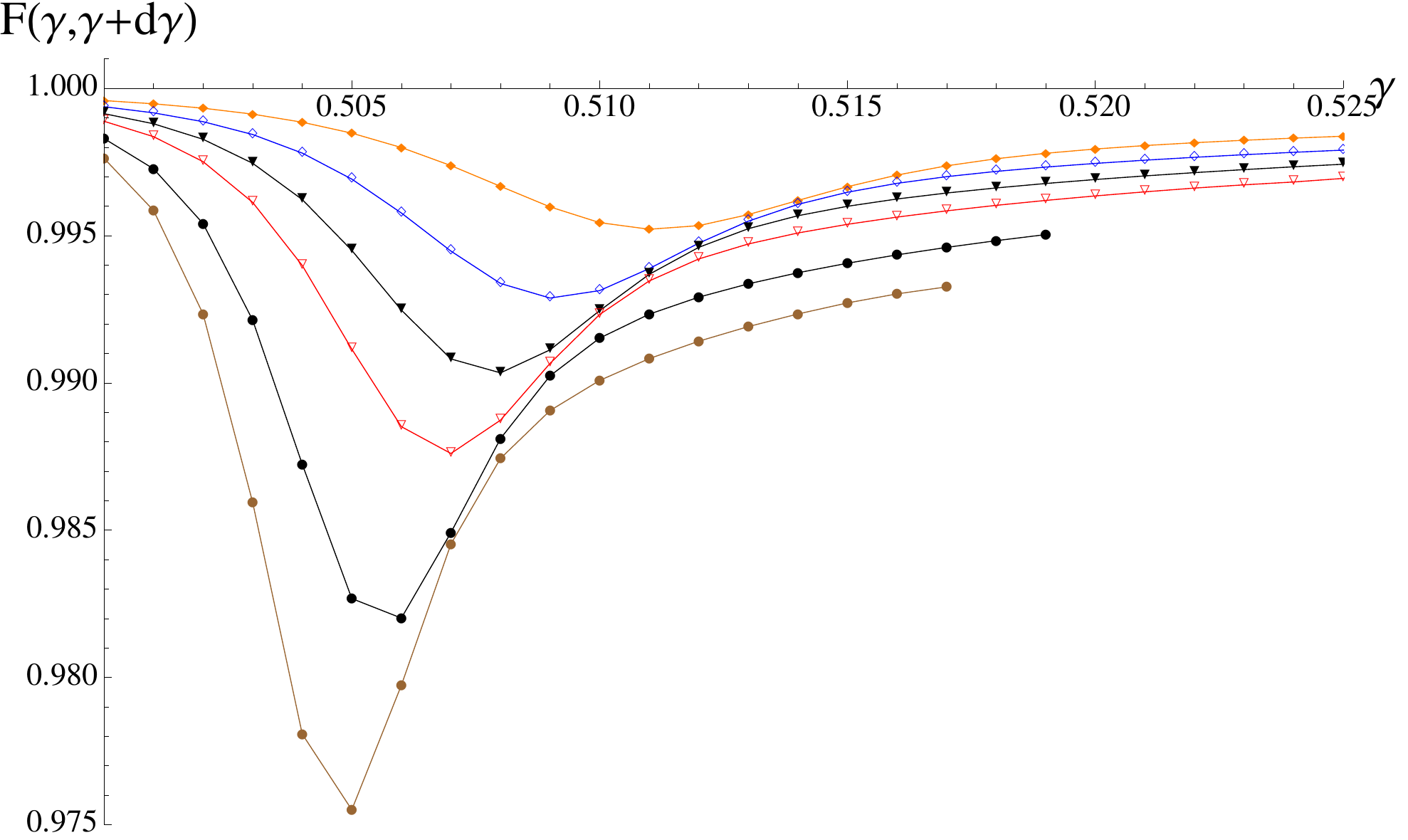}
\end{center}
\caption{\label{fig:2}Fidelity as a function of the coupling parameter. With $\gamma$ from $\gamma_{c}=0.5$ to $0.6$, $d\gamma=0.001$, $\omega_{0}=\omega=1$ and $\mathcal{N}=300$, $400$, $500$, $600$, $800$ and $1000$ (top to bottom). $N_{max}=8$.}
\end{figure}

\begin{figure}
\begin{center}
\includegraphics[scale=0.5]{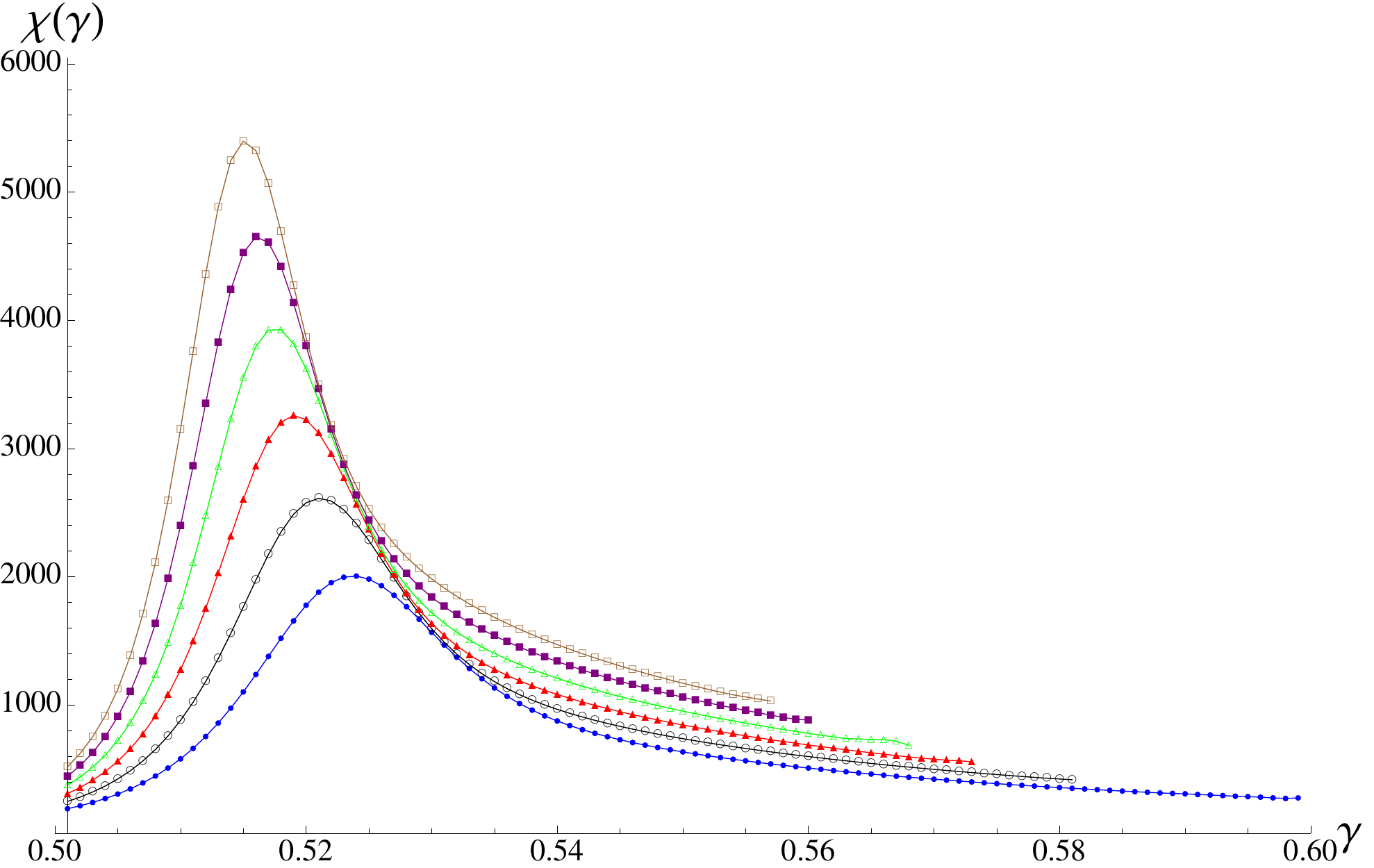}
\end{center}
\caption{\label{fig:3} Fidelity susceptibility   as a function of the coupling parameter. With $\gamma$ from $\gamma_{c}=0.5$ to $0.6$, $d\gamma=0.001$, $\omega_{0}=\omega=1$ and $\mathcal{N}=100$, $120$, $140$, $160$, $180$, $200$ (bottom to top). $N_{max}=8$.}
\end{figure}

\begin{figure}
\begin{center}
\includegraphics[scale=0.5]{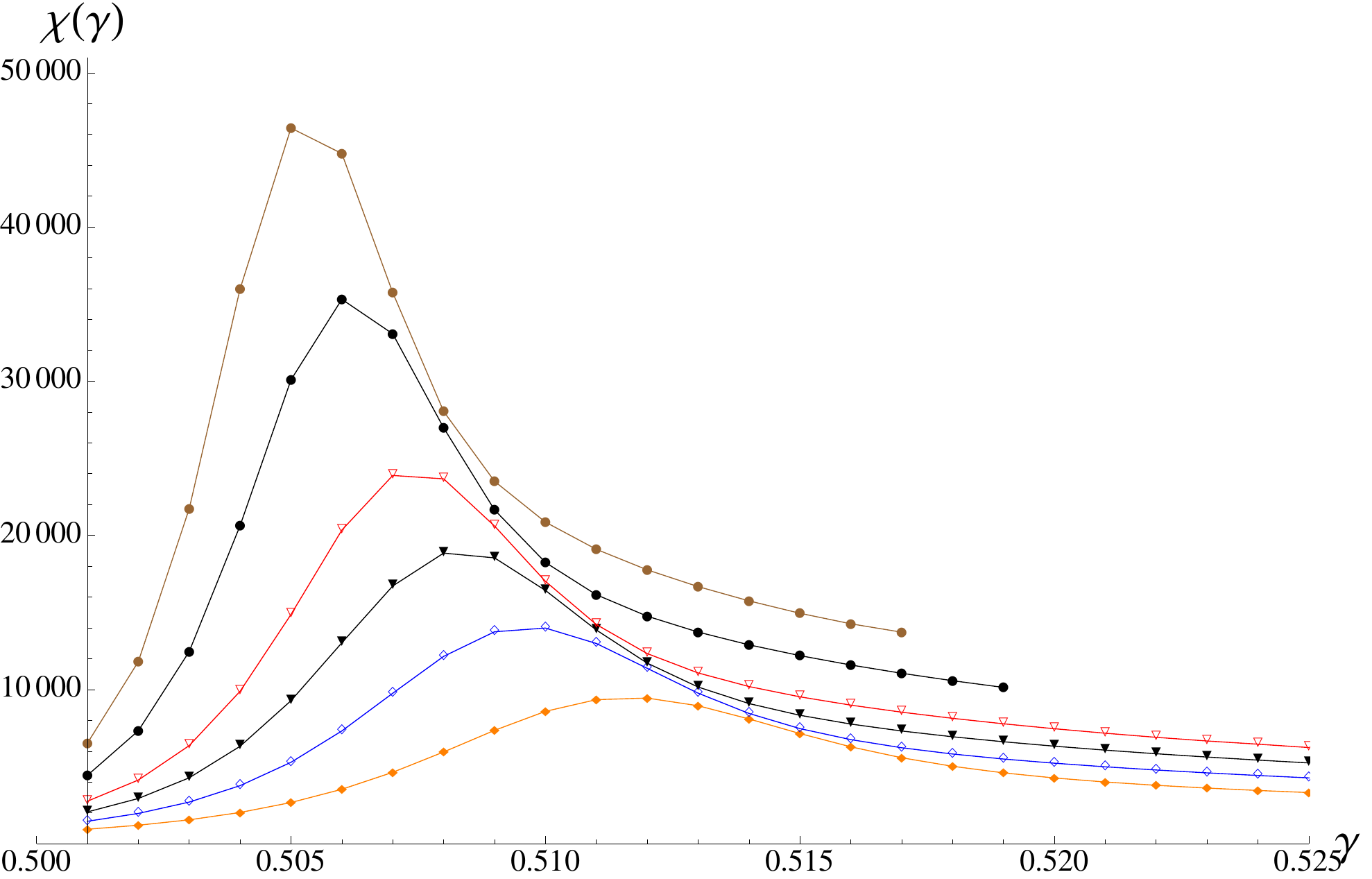}
\end{center}
\caption{\label{fig:4} Fidelity susceptibility   as a function of the coupling parameter. With $\gamma$ from $\gamma_{c}=0.5$ to $0.6$, $d\gamma=0.001$, $\omega_{0}=\omega=1$ and $\mathcal{N}=300$, $400$, $500$, $600$, $800$ and $1000$ (bottom to top). $N_{max}=8$.}
\end{figure}

We calculate the fidelity and its susceptibility as functions of the coupling $\gamma$ for the ground state by solving numerically the Hamiltonian. In figures \ref{fig:1} and \ref{fig:2} we show the fidelity for several values of $\mathcal{N}$ from $100$ to $1000$. The same goes for the fidelity susceptibility   in figures \ref{fig:3} and \ref{fig:4}. In these calculations we use $\omega=\omega_{0}=1$ (resonance) being $\gamma_{c}=0.5$ the critical value of the coupling in the thermodynamic limit. 



\begin{figure}
\begin{center}
\includegraphics[scale=0.5]{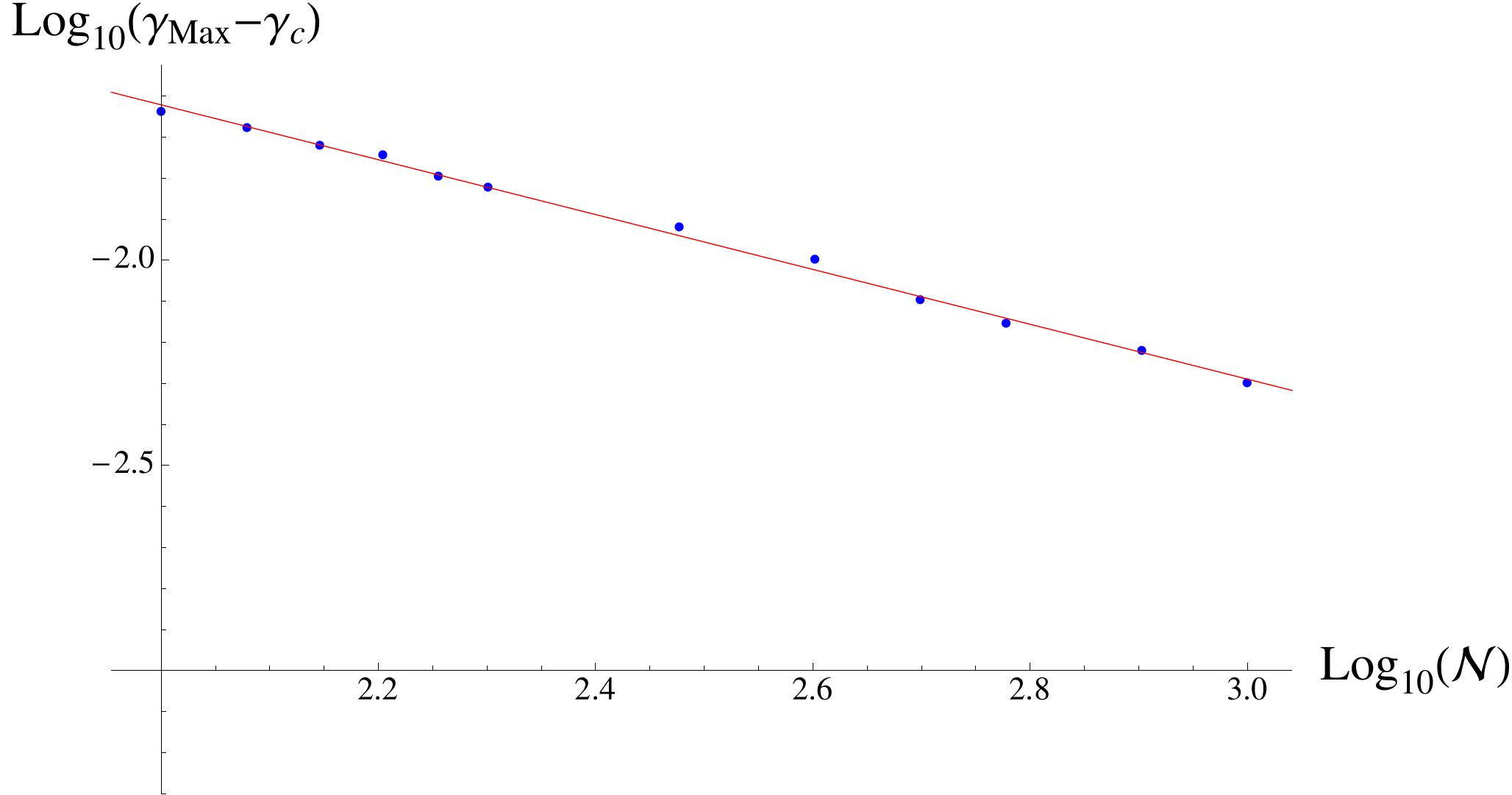}
\end{center}
\caption{\label{fig:5} $Log_{10}\left(\gamma_{max}\right)$ as a function of $Log_{10}\left(\mathcal{N}\right)$. With $\mathcal{N}$ from $100$ to $1000$. $\gamma_{c}=0.5$. The linear fit is shown in red.}
\end{figure}

We can locate the coupling's critical value  $\gamma_{max}$,  the value where the quantum phase crossover (the precursor of the QPT) takes place, by identifying the minimum of the fidelity and the maximum of its susceptibility. In figure \ref{fig:5} the value of $\gamma_{max}$ is shown for each $\mathcal{N}$ in a logarithmic scale. A linear fit gives us
\begin{equation}
\begin{split}
Log_{10}\left(\gamma_{max}-\gamma_{c}\right)&=-0.285094 - 0.668233 \, Log_{10}\left(\mathcal{N}\right) ,\\
\left(\gamma_{max}-\gamma_{c}\right)&=0.518688 \, \mathcal{N}^{-0.668223} .
\end{split}
\end{equation}
Where we can obtain the critical exponent $\nu=0.668223\simeq 2/3$, which agrees with previous results \cite{Ocasta12,Nahmad12}. 

\begin{figure}
\begin{center}
\includegraphics[scale=0.45]{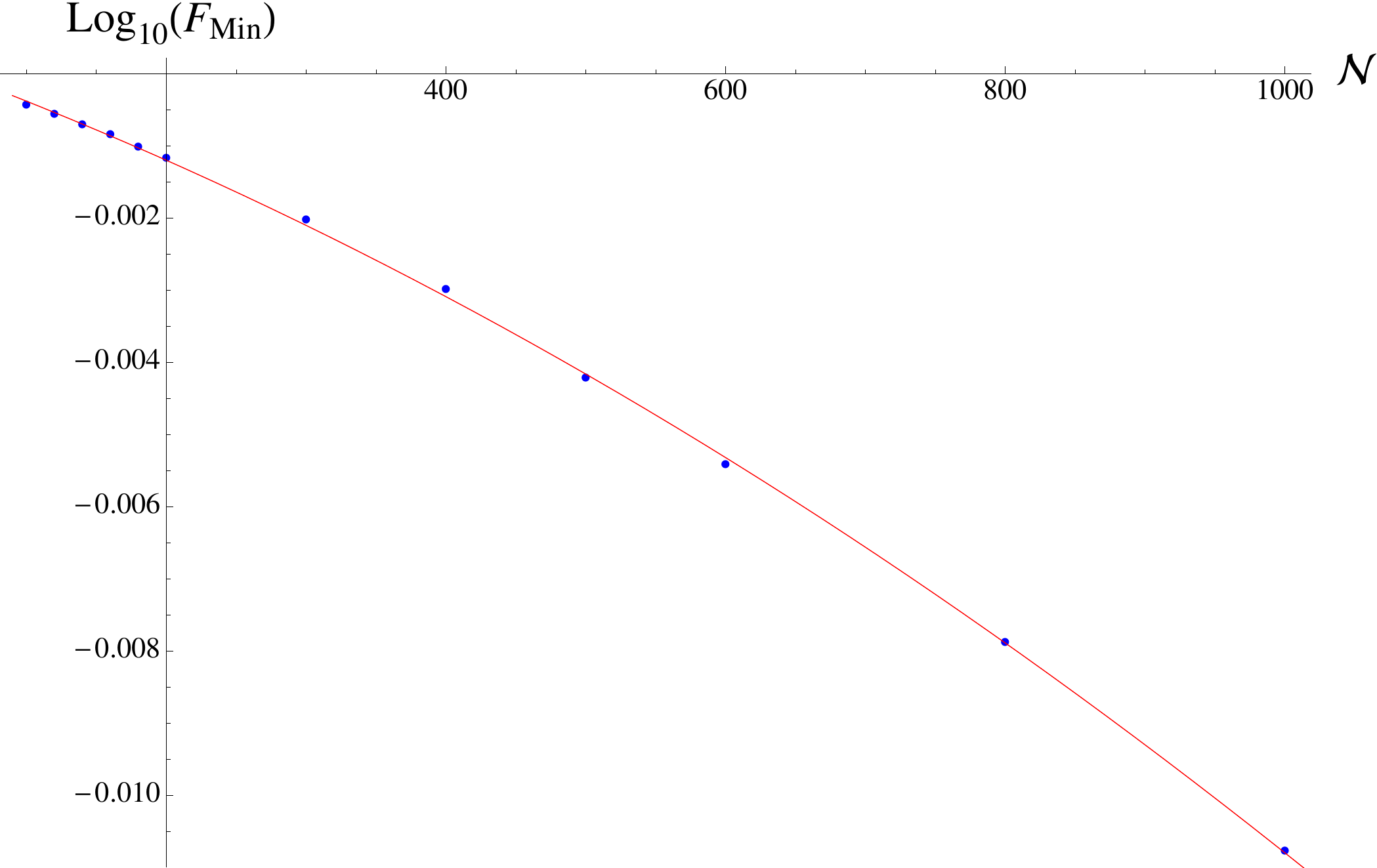}
\end{center}
\caption{\label{fig:6} $Log_{10}(F_{min)}$ as a function of $\mathcal{N}$. With $\mathcal{N}$ from $100$ to $1000$. $\gamma_{c}=0.5$. A quadratic fit is shown in red.}
\end{figure}

In figure \ref{fig:6} the logarithm of the minimum value of the fidelity $log_{10}\left(F_{min}\right)$ is plotted against the logarithm of the number of atoms. 
The points call for a quadratic fit, which is:
\begin{equation}
Log_{10}(F_{min})=0.000351536-6.90731\times 10^{-6} \mathcal{N}-4.23857\times 10^{-9} \mathcal{N}^{2} .
\end{equation}
We expect that, as we increase the number of atoms, the coefficient of the quadratic term will go to zero. In other words, the quadratic contribution is required by the small $\mathcal{N}$ values, from $100$ to $200$. 

\begin{figure}
\begin{center}
\includegraphics[scale=0.45]{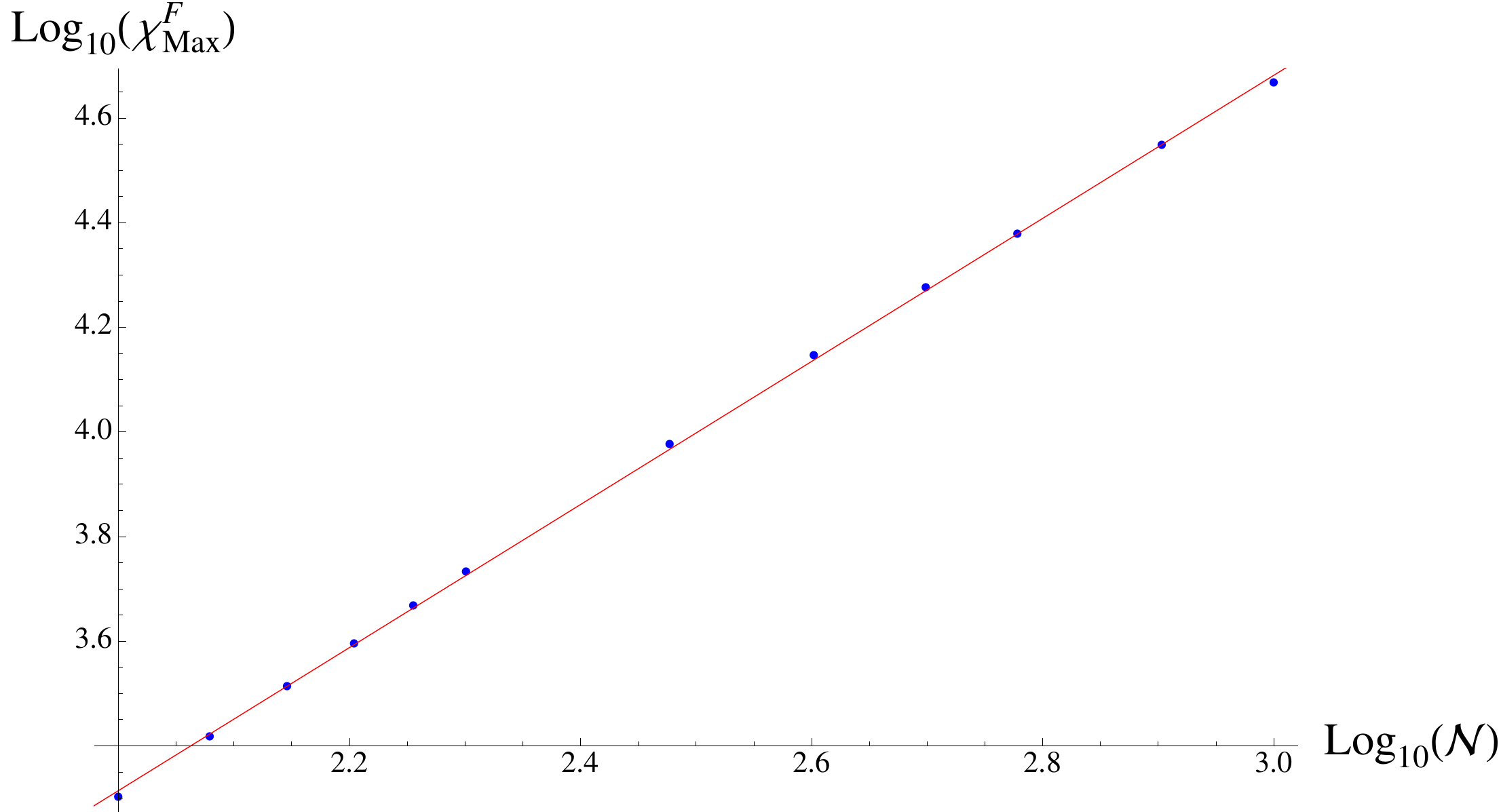}
\end{center}
\caption{\label{fig:7} $Log_{10}\left(\chi^{F}_{max}\right)$ as a function of $Log_{10}\left(\mathcal{N}\right)$. With $\mathcal{N}$ from $100$ to $1000$. $\gamma_{c}=0.5$. The linear fit is shown in red.}
\end{figure}

\begin{figure}
\begin{center}
\includegraphics[scale=0.55]{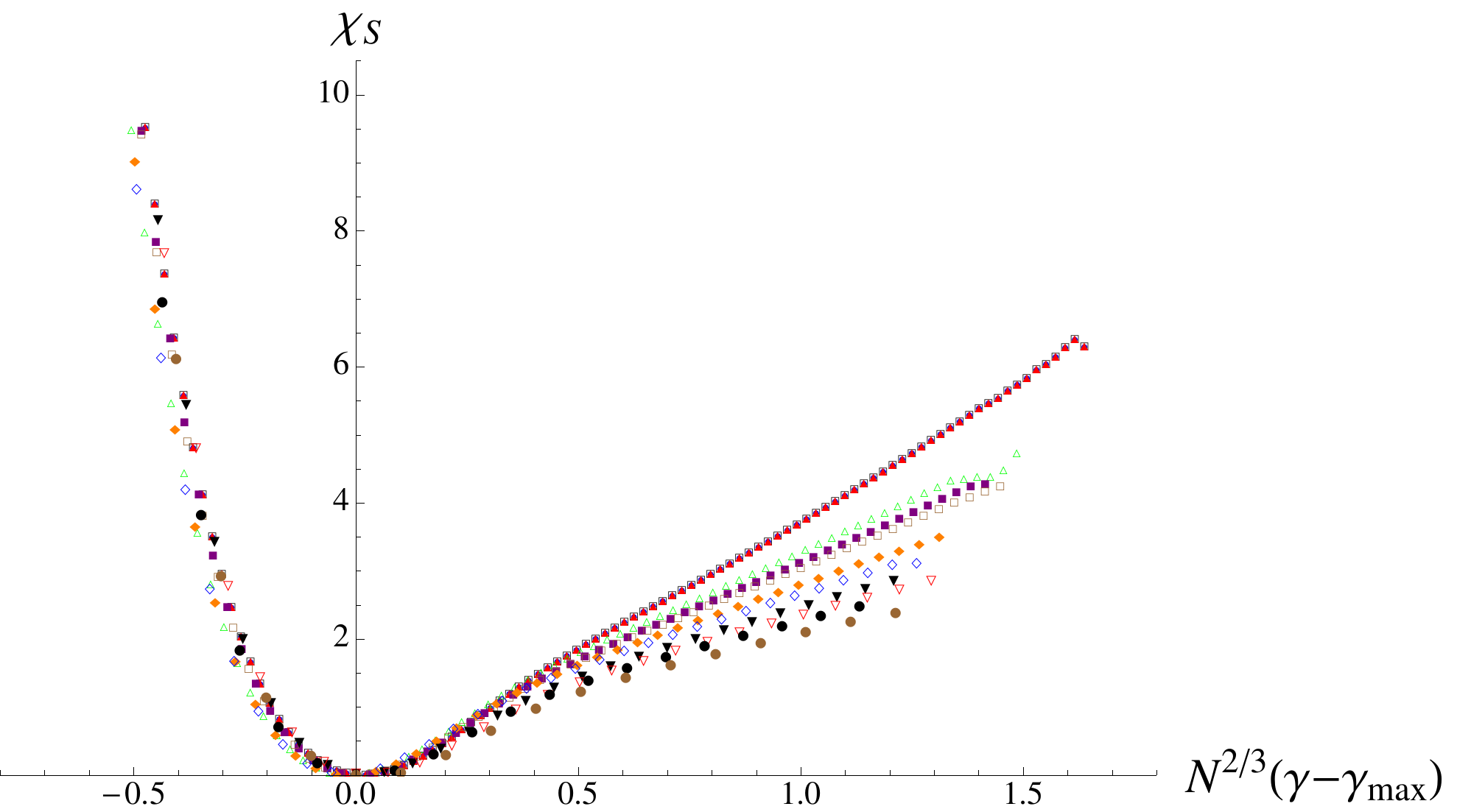}
\end{center}
\caption{\label{fig:8} Universal curve for $\chi_{S}$ as a function of $N^{\nu}\left(\gamma-\gamma_{max}\right)$. With $\mathcal{N}$ from $100$ to $1000$. $\gamma_{c}=0.5$, $\alpha=2/3$.}
\end{figure}

Fig. \ref{fig:7} displays the logarithm of maximum value of the fidelity susceptibility   $\chi^{F}_{max}$ as a function of the logarithm of the number of atoms. 
Fitting linearly the logarithmic curve between the maximum of the susceptibility and $\mathcal{N}$ we obtain:  
\begin{equation}
\begin{split}
Log_{10}\left(\chi^{F}_{max}\right)&=0.579291+ 1.36739 \, Log_{10}\left(\mathcal{N}\right),\\
\chi^{F}_{max}&=3.79569 \, \mathcal{N}^{1.36739} .
\end{split}
\end{equation}
The critical exponent is in this case $1.36739\simeq 4/3$ which agrees with the one found for the Lipkin-Meshkov-Glick model \cite{Hirsch13}, which belongs to the same universality class \cite{Dus04}.  
Also, we can calculate the universal curve of the specific susceptibility for every value of $\mathcal{N}$. We show the curve in figure \ref{fig:8}. The universal curve guarantees that the critical exponent is correct, because the curves for all $\mathcal{N}$ converge to one curve in the region around the critical value of the coupling strength $\gamma_{max}$. The results of figure \ref{fig:8} agree with the ones in \cite{Liu09}.

\begin{figure}
\begin{center}
\includegraphics[scale=0.5]{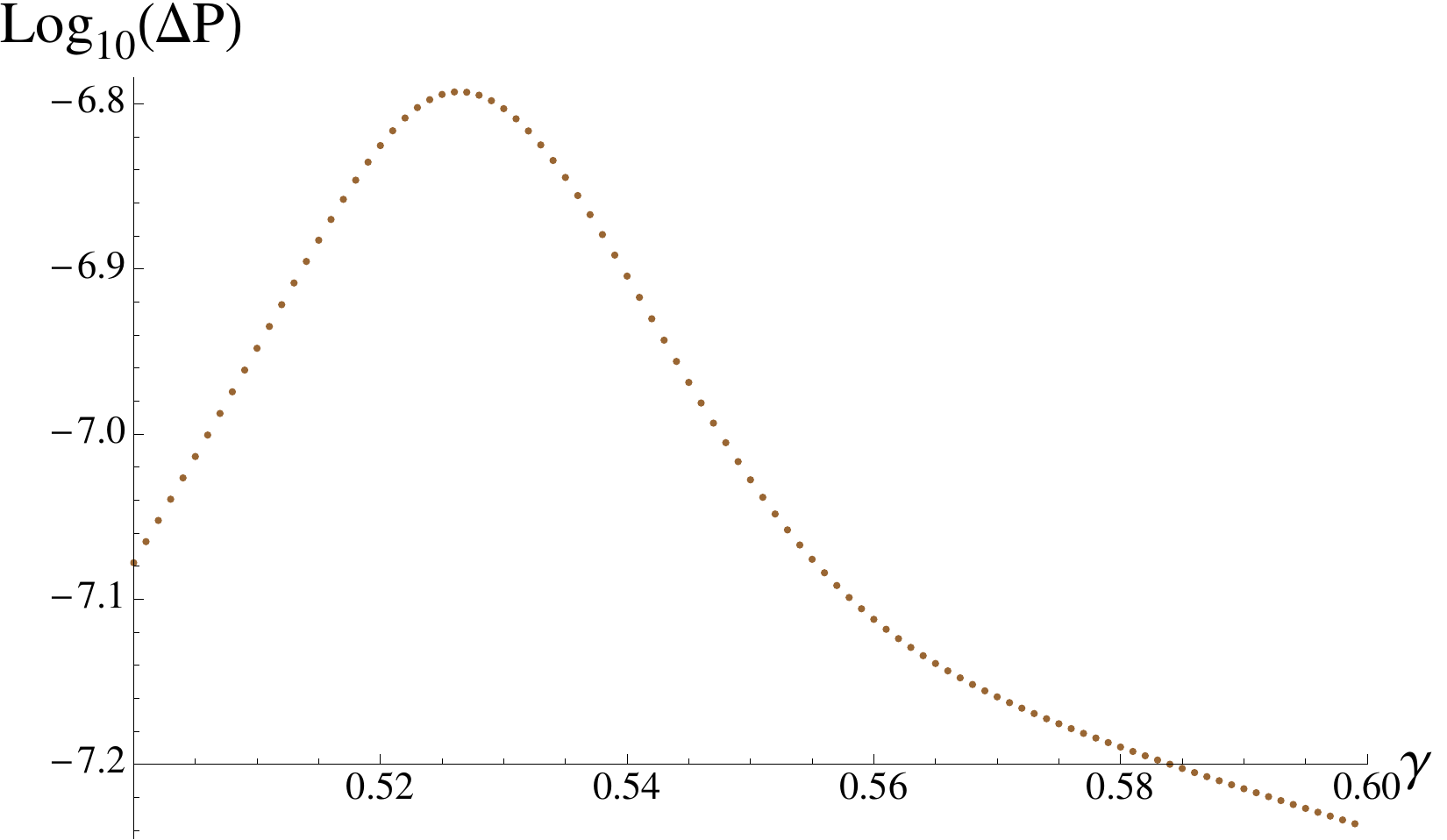}
\end{center}
\caption{\label{fig:9} $\Delta P$ for the ground state as a function of the coupling parameter. With $\gamma$ from $\gamma_{c}=0.5$ to $0.6$, $d\gamma=0.001$, $\omega_{0}=\omega=1$ and $\mathcal{N}=100$. $N_{max}=8$. In this example, the $\gamma_{max}=0.523$ and the value of the coupling where $\Delta P$ has its maximum is $\gamma=0.526$.}
\end{figure}

Finally, in figure \ref{fig:9} we show $\Delta P$ for the ground state as a function of the coupling for $\mathcal{N}=100$. As it can be observed, close to the phase transition precursor, which for this number of atoms takes place at $\gamma_{max}=0.523$, the numerical precision of the ground state wave function becomes smaller. The maximum of this curve occurs at a value of the coupling constant close to, but different from, the $\gamma_{max}$ calculated through the fidelity and its susceptibility. In all cases the $\Delta P$ is small enough to consider that the solution has converged. This behavior repeats for every $\mathcal{N}$. The maximum of the $\Delta P$ behaves in a similar way as $\gamma_{max}$ when the number of atoms grows, taking place closer and closer to $\gamma_c$ in the thermodynamic limit.  

%%%%%%%%%%%%%%%%%%%%%%%%%%%%%%%%%%%%%%%%%%%%%%%%%%

\section{Conclusions}
We have calculated the fidelity and its susceptibility for the ground state of the finite Dicke model, as functions of the coupling parameter strength, in resonance, for several values of the number of atoms. We located the phase transition for each value of $\mathcal{N}$ using the fidelity formalism, and characterized the phase transition by calculating the critical exponents of the critical values of the coupling and the maximum values of the susceptibility, by fitting logarithmically the curves of both as functions of $\mathcal{N}$. The critical exponents are
\begin{equation}
\left(\gamma_{max}-\gamma_{c}\right)\simeq N^{-0.668223}\simeq N^{-2/3}\,\,\,\mbox{and}\,\,\,\chi^{F}_{max}\simeq N^{1.36739}\sim N^{4/3}.
\end{equation}
Also, we fitted a quadratic curve of the logarithm of the fidelity as a function of the number of atoms. We validated the values found for the critical exponents plotting the universal curve of the specific susceptibility. Finally, we exhibited that the precision of the ground state wave function (which we use to determine the minimal truncation necessary to have the exact numerical solution) have a maximum near the finite $\mathcal{N}$ phase crossover. Interestingly, those maxima occur at a coupling values slightly different from the ones obtained through the maximum of the fidelity susceptibility  . 

\section{Acknowledgments}
This work was partially supported by CONACyT-M\'exico and DGAPA-UNAM project IN102811.

\section{Appendix}

In order to estimate the convergence in the wave function $|\Psi(N_{max})\rangle$, we assume that a similar diagonalization was performed with a truncation $N_{max}-1$, which provides $|\Psi(N_{max}-1)\rangle$. To compare both wave functions we extend the latter by assigning  $C_{N_{max},m}(N_{max}-1)=0$

We define the precision in the calculated wave function as:
\begin{equation}
\begin{split}
 \Delta P &\equiv 1-\left|\langle \Psi(N_{max}-1)|\Psi(N_{max})\rangle \right| \nonumber \\
&= 1- \left|\sum\limits_{N,N'=0}^{N_{max}}\sum\limits_{m,m'=-j}^j C_{N',m'}(N_{max}-1) C_{N,m}(N_{max}) \langle N';j,m' | N;j,m\rangle \right| \\
 &= 1- \left|\sum\limits_{N=0}^{N_{max}-1}\sum\limits_{m=-j}^j C_{N,m}(N_{max}-1) C_{N,m}(N_{max})\right| 
\end{split}
\end{equation}

We assume that $N_{max}-1$ is large enough to allow the wave function to be close to convergence. It implies that adding to the Hilbert space the states with $N_{max}$ photon excitations, the components $C_{N,m}, N\le N_{max}-1$ will have small changes, conserving their respective phases (but for a global one) with their magnitude remaining constant or slightly decreasing to allow for non-zero $C_{N_{max},m}$ new contributions. This condition can be expressed as
 
 \begin{equation}
 \left|C_{N,m}(N_{max}-1)\right| \ge \left|C_{N,m}(N_{max})\right| , \,\,\, N\le N_{max}-1.  \nonumber
 \end{equation}

It follows that 
  \begin{equation}
\begin{split}
\left|\sum\limits_{N=0}^{N_{max}-1}\sum\limits_{m=-j}^j C_{N,m}(N_{max}-1) C_{N,m}(N_{max})\right| \ge
 \sum\limits_{N=0}^{N_{max}-1}\sum\limits_{m=-j}^j \left|C_{N,m}(N_{max})\right|^2  \nonumber  \end{split},
 \end{equation}
and
 \begin{equation}
\begin{split}
 \Delta P  &\leq 1- \sum\limits_{N=0}^{N_{max}-1}\sum\limits_{m=-j}^j \left|C_{N,m}(N_{max})\right|^2  \nonumber \\
 &= \sum\limits_{m=-j}^j \left|C_{N_{max},m}(N_{max})\right|^2 . \nonumber
 \end{split}
 \end{equation}

We employ the equality to obtain an upper bound to the precision of the calculated wave functions.

%%%%%%%%%%%%%%%%%%%%%%%%%%%%%%%%%%%%%%%%%%%%%%%%%%

%%%%%%%%%%%%%%%%%%%%%%%%%%%%%%%%%%%%%%%%%%%%%%%%%%
%%%%%%%%%%%%%%%%%%%%%%%%%%%%%%%%%%%%%%%%%%%%%%%%%%

\end{document}